\documentclass[10pt]{iopart}

\usepackage[cp1251]{inputenc}
\usepackage{iopams}
\usepackage{graphicx,epsfig}
\begin{document}

\title[Dirac theory]{Dirac theory as a one-particle relativistic quantum mechanics in the space of unit two-component spinors}

\author{N. L. Chuprikov}

\address{Tomsk State Pedagogical University, 634041, Tomsk, Russia}
\ead{chnl@tspu.edu.ru} \vspace{10pt}

\begin{abstract}
Using the example of a Dirac particle in external static fields, Dirac theory is formulated as a single-particle quantum
theory in the space of normalized two-component spinors. In this formulation, the Dirac operator is ``split'' into two
two-component operators: one bounded below (in the nonrelativistic limit, it coincides with the Pauli operator), and the
other bounded above. In the two-component formulation, exact analytical expressions are obtained for all observables,
including the Dirac Hamiltonian, which are then investigated in the nonrelativistic and ultrarelativistic limits. A
general solution to the equation for the unit two-component spinor describing a free particle is presented. The proposed
formulation is compared with the original Foldy-Wouthuysen (FW) formalism, as well as with the exact FW formalism based on
the ``wave function reduction condition''.

\end{abstract}

\newcommand{\ppp}{\mbox{\hspace{5mm}}}
\newcommand{\ooo}{\mbox{\hspace{3mm}}}
\newcommand{\ooa}{\mbox{\hspace{1mm}}}
\newcommand{\ppd}{\mbox{\hspace{18mm}}}
\newcommand{\ppt}{\mbox{\hspace{34mm}}}
\newcommand{\ppo}{\mbox{\hspace{10mm}}}
\newcommand{\lcom}{\lambda\hspace{-1mm}\bar{}\hspace{1mm}}

\section{Introduction}

As is well known, in deriving his equation, Dirac sought to construct a theory describing the quantum dynamics of a
relativistic electron that would be both relativistically invariant and satisfy the fundamental requirements of quantum
mechanics. Dirac succeeded in constructing a relativistically invariant theory, which, as it turns out, not only explains
the presence of spin in the electron but also the existence of the positron -- the electron's antiparticle. However,
seeing the quantum mechanics of individual particles in this theory proved so difficult that the problem remains unsolved
to this day (see, for example, \cite{Thal,Grei,Wat}). First of all, this concerns the basic quantum mechanical concepts of
`state', `observable' and `Hamiltonian', which in the Dirac theory acquire properties that are not characteristic of
single-particle quantum mechanics.

In particular, there is still no generally accepted answer to the question of why the number of functions required to
describe the electron's state doubles when moving from Pauli to Dirac theory. There is still no clear answer to the
question of whether operators, acting in the space of four-component Dirac bispinors, can be considered operators of
observables. And, of course, the most questionable is the Dirac Hamiltonian, which is unbounded not only above but also
below. On the one hand, thanks to this property, Dirac's theory predicted the existence of positrons, as well as quantum
mechanical phenomena accompanied by the creation and annihilation of electron-positron pairs. However, on the other hand,
this property means that Dirac's theory lacks a ground (stable) single-particle state, which contradicts one of the
fundamental principles of quantum mechanics.

As a result, a widely held view has emerged that a consistent formulation of the Dirac theory as single-particle quantum
mechanics is fundamentally impossible, and that the Dirac equation should be considered as a classical field theory
equation requiring quantization (see, for example, \cite{Ber,Sch}). As is streesed in \cite{Bogo}, ``{\it the introduction
of classical fields into physics is dictated by reasons of relativistic invariance}.''

However, relativistic invariance does not preclude the construction of ``non-field'' formulations of the Dirac
relativistic theory. It is necessary to take into account the fact that there are processes involving individual
relativistic electrons that are not accompanied by multiparticle effects. Although a consistent ``single-particle''
description of these processes has been developed within the framework of quantum field theory (see \cite{Git}), the
question of constructing a single-particle relativistic quantum theory without using the concept of an ``electron field''
remains a pressing problem. This is primarily necessary to justify non-relativistic quantum mechanics, since the boundary
between the quantum dynamics of a free non-relativistic particle and a free relativistic particle is relative -- a free
particle that is non-relativistic in one inertial frame of reference is relativistic in another; and vice versa.

Since the concept of a wave function describing the state of a particle plays a key role in quantum mechanics, to
construct a ``non-field'' single-particle formulation of Dirac theory, one must first answer two interrelated questions:
``Are the components of the Dirac bispinor, like the components of the Pauli spinor, wave functions?'' and ``What explains
the doubling of the number of functions required to describe the state of an electron when going from the Pauli equation
to the Dirac equation?''

For many years, the main candidate for a single-particle formulation was the Dirac theory in the Foldy-Wouthuysen (FW)
representation. Its main idea is expressed in the following arguments (see p. 98 in \cite{Sch}): ``A free Dirac particle
with positive energy has two independent states for each value of momentum. These correspond to two possible spin
directions. According to the principles of quantum mechanics, each such pair of physical states must be represented by
exactly two vectors in Hilbert space. Therefore, the usual formulation of Dirac theory contains redundancy in the
representation of these vectors, since the corresponding wave functions have four components.''

As is seen, this reasoning is based on the assumption that each component of the Dirac bispinor is a quantum-mechanical
wave function, and the doubling of the number of unknown functions in the transition from the Pauli equation to the Dirac
equation is due to the appearance of states with negative energies. This is why Foldy and Wouthuizen set themselves the
goal of finding ``\ldots a representation of the Dirac theory in which, for both relativistic and non-relativistic
energies, positive-energy states and negative-energy states are separately described by two-component wave functions''
(see p. 30 in \cite{Fol}). They found a transformation in which the transformed Dirac operator has a block-diagonal form;
it does not contain odd operators mixing states with positive and negative energies. For a free particle, this
transformation was found in closed form. For a particle in external fields, this method was developed as an approximate
approach that yields a block-diagonal relativistic Hamiltonian at any step of the expansion in powers of the parameter
$1/m$.

By now, approaches have been developed (see, for example, \cite{Bar,Blou,Rei1,Rei2}) that extend the applicability of the
original stepwise FW transformation to the case of arbitrary external fields. In addition, an exact version of the FW
transformation was developed for an arbitrary vector potential \cite{Cas,Eric,Eri,Oli}, for a scalar potential commuting
with an odd operator \cite{Sil0}, and for the special case of a $\mathbf{r}$-depending scalar potential \cite{Niki}. As
for an arbitrary $\mathbf{r}$-depending scalar potential, for this case the exact FW Hamiltonian was found in \cite{Nez1}
within the framework of a new version \cite{Nez} of the exact FW transformation, which, in our opinion, differs
fundamentally from the previous version and will be considered separately.

The main merit of the FW transformation is that it ``...provides the best possibility of obtaining a meaningful classical
limit of the relativistic quantum mechanics'' \cite{Nez}. However, there are reasons to doubt the basic idea of the FW
transformation, according to which the four-component Dirac bispinor is a four-component quantum mechanical wave function
and, as a consequence, ``the usual formulation of Dirac theory contains redundancy in the representation of the vectors
describing the state of a relativistic electron''.

By Roger Penrose, there is no `redundancy' in the Dirac bispinor. On page 624 in \cite{Pen} he writes: ``{\it In fact, the
particle described by the Dirac equation has only 2 spin components, even though the wave function has 4 components.
Mathematically, the reason for this is closely related to the fact that the Dirac equation
$\not\hspace{-0.5mm}\partial\psi=-iM\psi$ is a first-order equation, and its solution space is covered by only half the
solutions compared to the case of the second-order wave equation $({\not\hspace{-0.5mm}\partial}^2 + M^2)\psi=0$. (This
equation is also satisfied by the ``anti-Dirac'' equation $\not\hspace{-0.5mm}\partial\psi=+iM\psi$, which is the Dirac
equation for a negative rest mass $M$.) Physically, this ``counting'' of solutions of the Dirac equation must take into
account the fact that the degrees of freedom of the electron's antiparticle, namely the positron, are also hidden in the
solutions of the Dirac equation. However, it would be a mistake to think that two components of the Dirac equation refer
to the electron and the other two to the positron.}''

We share this viewpoint and add that all four components of the Dirac bispinor correspond to electron states with positive
energy (or all four correspond to electron states with negative energy, which are transformed into positron states with
positive energy when charge conjugation is taken into account). The key point is that, contrary to the Foldy-Wouthuysen
idea, the components of the Dirac bispinor are not wave functions. To describe the state of a relativistic particle with a
given spin projection and a given energy sign, two components of the Dirac bispinor are required. To paraphrase Bogolyubov
and Shirkov's words above, the doubling of the number of functions required to describe the state of a free electron upon
the transition from Pauli theory to Dirac theory ``...is dictated by reasons of relativistic invariance''.

Thus, the very first statement in the abstract of the article \cite{Fol} ``{\it By a canonical transformation on the Dirac
Hamiltonian for a free particle, a representation of the Dirac theory is obtained in which positive and negative energy
states are separately represented by two-component wave functions}'' misleads, becase all four components of the Dirac
bispinor for a free particle correspond to states with {\it positive} energy, and there is no need to separate these
components.

Of importance is also to stress that the FW transformation contains a non-analytic function of the momentum operator and,
as a consequence, is fundamentally different from the unitary transformation linking the Dirac and Weyl representations.
Mathematically, the unitary equivalence of these two representations is manifested in the fact that the Dirac bispinor in
the Weyl representation, as in the standard representation, obeys a system of four first-order differential equations. At
the same time, for a free nonrelativistic particle, the FW representation yields two second-order differential equations
for positive-energy states and two second-order differential equations for negative-energy states. That is, unlike the
Weyl representation, the FW representation is not mathematically equivalent to the standard representation of the Dirac
theory (see also \cite{Bir,Matz}).

The definition of observables' operators in the FW representation also raises questions. The point is that within the FW
approach, an arbitrary operator in the standard representation is transformed into a four-component matrix operator, which
does not have a block-diagonal form like the FW Hamiltonian. This means that the FW transformation does not actually
separate the states of the Dirac particle into states with positive and negative energy (these subspaces are not invariant
under the action of the operators of all observables).

To avoid this situation, Foldy and Wouthuysen assume in \cite{Fol} (see also \cite{Sil}) that the operators of observables
in the Dirac representation have no physical meaning and redefine these observables so that they have a block-diagonal
form in the FW representation. As a result, the FW-operators ``mean position'', ``mean velocity'', ``mean orbital angular
momentum'', and ``mean spin angular momentum'' emerge.


The very definition of the FW-operators (including the ``mean position operator'') is logically inconsistent. But there is
another reason why the definition of the ``mean position operator'' in the FW representation is objectionable. This is
because the projections of the mean position operator commute with each other. At the same time, the classical equations
formulated for the relativistic electron in the Frenkel-Bargmann-Michel-Telegdi approach imply (see \cite{Der}) that the
projections of the particle coordinate must be represented in relativistic quantum mechanics by non-commuting operators.

A special place in the formalism of the FW transformation is occupied by the paper \cite{Nez}, where it is introduced the
so-called ``wave function reduction condition'', which ``...{\it consists in the nullification of either upper or lower
components of the bispinor wave function}''. According to this condition, only those Dirac bispinors can undergo the FW
transformation, for which the corresponding transformed bispinors contain only two upper (or two lower) nonzero
components. However, the true meaning of this condition can only be understood in the context of the approach proposed
here, and we will dwell on it in Section \ref{Nezn}.


In the present paper, we develop an approach according to which the ``large'' and ``small'' components of a unit-norm
Dirac bispinor, satisfying the Dirac equation (a system of first-order differential equations), either both correspond to
a state with positive energy, or both correspond to a state with negative energy. As will be shown in sections
\ref{elimination} and \ref{equ}, there exists a unit two-component spinor $f$ (satisfying a system of second-order
differential equations (\ref{3f})), to which both of these components are uniquely related (see (\ref{888})). The Dirac
operator with vector potential $\mathbf{A}(\mathbf{r})$ and scalar potential $V(\mathbf{r})$ ``splits'' into two
``two-component'' Hamiltonians acting in the space of spinors $f$: one bounded below is a relativistic analog of the Pauli
Hamiltonian; the other bounded above (both operators are presented in closed form).

For one-particle states with an energy bounded above, in addition to the two-component Hamiltonian, we define (in closed
form) two-component analogues of the operators of other observables (see Section \ref{operators}) and then investigate
them in the nonrelativistic (Section \ref{nonrel}) and ultrarelativistic limits (Section \ref{ultrarel}). That is, in
fact, we formulate the Dirac theory as a two-component single-particle relativistic quantum mechanics (provided that the
vector and scalar potentials do not lead to the creation and annihilation of electron-positron pairs). As shown, the
projections of the particle coordinate in this formulation are represented by two-component operators commuting with each
other only in the strict non-relativistic limit.

It is important to emphasize that our approach, although it denies the basic Foldy-Wouthuysen's idea, according to which
the components of the Dirac bispinor simultaneously describe states with positive and negative energy, nevertheless
confirms the correctness of the expressions for a transformed Hamiltonian obtained in closed form for a free particle
\cite{Fol}, for a particle in an external vector potential \cite{Cas,Eric}, and for a particle in an external scalar
potential commuting with an odd operator \cite{Sil}.

As will be shown in Section \ref{Nezn}, the ``wave function reduction condition'' introduced in \cite{Nez} selects only
those Dirac bispinors for which all four components correspond to either only positive or only negative energy. For a
particle in a static scalar potential depending on $\mathbf{r}$, a comparison of our approach with the exact FW
transformation \cite{Nez1} based on this condition reveals an interesting relationship between these approaches.

\section{The large and small components of the unit Dirac bispinor as ``projections''
of the unit two-component spinor}\label{elimination}

Dirac showed that the operator $i\hbar\partial/\partial t$ acts in the space of normalized bispinors in the standard
representation as the operator $H_D$:
\begin{eqnarray} \label{1}
\fl i\hbar\frac{\partial{\Psi}}{\partial t}=H_D\Psi;\ppp H_D=\left(
\begin{array}{cc}
V+mc^2 & c\mathcal{P} \\
c\mathcal{P} & V-mc^2
\end{array} \right) \equiv\left(
\begin{array}{cc}
H_{11} & H_{12} \\
H_{21} & H_{22}
\end{array} \right),
\end{eqnarray}
where $\mathcal{P}=(\vec{\sigma}\vec{\pi})$; $\vec{\sigma}=(\sigma_x,\sigma_y,\sigma_z)$; $\sigma_x$, $\sigma_y$ and
$\sigma_z$ is the Pauli matrix; ${\vec{\pi}}={\vec{p}}-\frac{e}{c}\vec{A}$, ${\vec{p}}=-i\hbar \vec{\nabla}$,
$V=e\varphi$. Bispinors $\Psi$ belong to the Hilbert space
\begin{eqnarray*}
\fl \mathcal{H}_4=L^2(\mathbf{R}^3)\oplus L^2(\mathbf{R}^3)\oplus L^2(\mathbf{R}^3)\oplus L^2(\mathbf{R}^3)
\end{eqnarray*}
with the scalar product and normalization rule for arbitrary bispinors $\Psi=(\Phi,\chi)^\top$ and
$\Psi_1=(\Phi_1,\chi_1)^\top$:
\begin{eqnarray*}
\fl \langle \Psi_1|\Psi\rangle=\int_{\mathbf{R}^3} \left(\Phi_1^\dag\Phi+\chi_1^\dag\chi\right)d^3\mathbf{r},\ppp \langle
\Psi|\Psi\rangle\equiv\int_{\mathbf{R}^3} \left(\Phi^\dag\Phi+\chi^\dag\chi\right)d^3\mathbf{r}=1.
\end{eqnarray*}
We will assume that the domain of $H_D$, as an eesentially self-adjoint operator, is
\begin{eqnarray*}
\fl \mathcal{D}(H_D)=\mathcal{S}_4=S(\mathbf{R}^3)\oplus S(\mathbf{R}^3)\oplus S(\mathbf{R}^3)\oplus S(\mathbf{R}^3),
\end{eqnarray*}
where $\mathcal{S}$ is the Schwartz space. Our goal is to find out how the operator $i\hbar\partial/\partial t$ acts in
the space of two-component spinors with unit norm, and to establish a connection between Dirac theory and non-relativistic
Pauli theory.

Let us write the Dirac equation as a system of equations for the large and small components
\begin{eqnarray} \label{3}
\fl i\hbar\frac{\partial \Phi}{\partial t}=(V+mc^2) \Phi+c \mathcal{P} \chi; \ppp i\hbar\frac{\partial \chi}{\partial t}=c
\mathcal{P} \Phi+(V-mc^2) \chi
\end{eqnarray}
and suppose that for any bispinor $\Psi$ there always exists a unit two-component spinor $f$ such that
\begin{eqnarray} \label{888}
\fl \Phi=\cos(\phi)f,\ppp \chi=\sin(\phi)f,
\end{eqnarray}
where $\phi$ is assumed to be a Hermitian operator (obviously, $\Phi^\dag\Phi+\chi^\dag\chi=f^\dag f$).

Substituting these spinors $\Phi$ and $\chi$ into equations (\ref{3}), we obtain
\begin{eqnarray} \label{1s}
\fl \left[\left(D_t-V-mc^2\right)\cos(\phi) -c\mathcal{P}\sin(\phi)\right]f=0,\nonumber\\
\fl \left[\left(D_t-V+mc^2\right)\sin(\phi)-c\mathcal{P}\cos(\phi)\right] f=0,
\end{eqnarray}
where $D_t= i\hbar\partial/\partial t$. Since the spinor $f$ in these equalities is arbitrary, the following compatibility
conditions must be satisfied:
\begin{eqnarray} \label{1t}
\fl \sin\phi=(c\mathcal{P})^{-1}(D_t-V-mc^2)\cos\phi=(D_t-V+mc^2)^{-1}c\mathcal{P}\cos\phi.
\end{eqnarray}

The right equality in (\ref{1t}) is an equation for the operator $D_t$ in the space of two-component spinors:
\begin{eqnarray} \label{2t}
\fl (c\mathcal{P})^{-1}(D_t-V-mc^2)=(D_t-V+mc^2)^{-1}c\mathcal{P}.
\end{eqnarray}
In our approach it plays the key role in the procedure of splitting the original Dirac operators into two two-component
semibounded relativistic Hamiltonians. And at first glance, as in the FW approaches, the presence of a scalar potential
$V$ that does not commute with the operator $\mathcal{P}$ should complicate this procedure.

However, this is not the case. As follows from Eq. (\ref{2t}), if we take the operator $X=D_t-V$ as the unknown operator,
then this equation takes the form
\begin{eqnarray*}
\fl (c\mathcal{P})^{-1}(X-mc^2)=(X+mc^2)^{-1}c\mathcal{P}.
\end{eqnarray*}
This means that the desired operator $X$ commutes with $\mathcal{P}$, and as a result we obtain two roots of Eq.
(\ref{2t}):
\begin{eqnarray} \label{3t}
\fl D_t^{(\pm)}=X^{(\pm)}+V=V\pm\sqrt{m^2c^4+(c\mathcal{P})^2}.
\end{eqnarray}
The first one obviously defines an operator $D_t$ bounded below, and the second one defines an operator $D_t$ bounded
above. Now we have to define the operator $\phi$ for each of these two roots.

From condition (\ref{1t}) it follows that
\begin{eqnarray*}
\fl \tan^2\phi=\frac{D_t-V-mc^2}{D_t-V+mc^2}.
\end{eqnarray*}
Thus, for the root $D_t^{(+)}$ we have $\tan^2\phi=\mu/M$, where
\begin{eqnarray*}
\fl M=\frac{D_t^{(+)}-V+mc^2}{2mc^2}=\frac{m}{2}\left[\sqrt{1+\xi^2}+1\right], \ooa
\mu=\frac{D_t^{(+)}-V-mc^2}{2mc^2}=\frac{m}{2}\left[\sqrt{1+\xi^2}-1\right]
\end{eqnarray*}
where $\xi=\frac{\mathcal{P}}{mc}$. Or, more exactly, in accordance with (\ref{1t}), for this root
\begin{eqnarray}\label{fi1}
\fl \sin\phi=\frac{\mathcal{P}}{|\mathcal{P}|}\sqrt{\frac{\mu}{M+\mu}},\ppp \cos\phi=\sqrt{\frac{M}{M+\mu}},
\end{eqnarray}
where $\mathcal{P}^2= 4\mu M c^2$. Similarly, it is easy to show that the root $D_t^{(-)}$ leads to the angle operator
$\phi_{(-)}=\phi+\pi/2$:
\begin{eqnarray}\label{fi2}
\fl \sin\phi_{(-)}=\cos\phi,\ppp \cos\phi_{(-)}=-\sin\phi.
\end{eqnarray}
In this case, the mass operators $M$ and $\mu$ are defined by the expressions
\begin{eqnarray} \label{10t}
\fl M_{(-)}=\frac{D_t^{(-)}-V+mc^2}{2mc^2}=-\mu,\ppp \mu_{(-)}=\frac{D_t^{(-)}-V-mc^2}{2mc^2}=-M.
\end{eqnarray}

As we emphasized above, the roots $D_t^{(+)}$ and $D_t^{(-)}$ are semi-bounded operators. However, they are not the
searched-for two-component analogues of the four-component Dirac operator. This is because they act on the spinors
$\cos\phi f$ and $\sin\phi f$, rather than directly on the unit spinor $f$. These roots do not guarantee the equality of
the average energy of the particle in the four-component and two-component formulations of Dirac theory, for each of these
two ranges of the particle's energy. In this connection, our next step is to find for both roots the corresponding
self-adjoint two-component analogues of the Dirac operator and the equation for the unit spinor $f$.

\section{Dirac equation in the space of unit two-component spinors} \label{equ}

For solving this problem, we consider the Dirac equation (\ref {3}) not as an equation for the bispinor $\Psi$, but as a
condition according to which the operator $i\hbar\partial/\partial t$ acts in the space of arbitrary unit bispinors $\Psi$
as the operator $H_D$. We multiply both sides of this condition by the arbitrary unit bispinor $\Psi_1$, integrate over
$\mathbf{R}^3$ and take into account the fact that the operator $H_D$ is self-adjoint. As a result, we obtain two
equalities
\begin{eqnarray}\label{12}
\fl i\hbar \left\langle\Psi_1\bigg|\frac{\partial \Psi}{\partial t}\right\rangle=\langle\Psi_1|H_D\Psi\rangle= \langle
H_D\Psi_1|\Psi\rangle
\end{eqnarray}
which must be valid for arbitrary bispinors $\Psi_1$ and $\Psi$.

Let us choose the root $D_t^{(+)}$ and write both bispinors in terms of arbitrary spinors $f_1$ and $f$, using relations
(\ref{888}):
\begin{eqnarray}\label{1q}
\fl \Psi=\left(
\begin{array}{cc}
\cos\phi \\
\sin\phi
\end{array} \right)f,\ppp \Psi_1=\left(
\begin{array}{cc}
\cos\phi \\
\sin\phi
\end{array} \right)f_1.
\end{eqnarray}
Substituting these expressions into (\ref{12}), it is easy to show that both equalities are reduced to the form
\begin{eqnarray}\label{888a}
\fl i\hbar \left\langle f_1\bigg|\frac{\partial f}{\partial t}\right\rangle= \langle f_1| H_{(+)} f\rangle = \langle
H_{(+)}f_1|f\rangle,
\end{eqnarray}
where
\begin{eqnarray}\label{888b}
\fl H_{(+)}=\cos\phi\ooa H_{11}\cos\phi+\cos\phi\ooa H_{12}\sin\phi+\sin\phi\ooa H_{21}\cos\phi+\sin\phi\ooa
H_{22}\sin\phi.
\end{eqnarray}
Since the spinors $f_1$ and $f$ are arbitrary, it follows from the second equality in (\ref{888a}) that the operator
$H_{(+)}$ is just the searched-for self-adjoint two-component analogue of the Dirac operator, and from the first it
follows that the spinor $f$ must satisfy the equation
\begin{eqnarray} \label{3f}
\fl i\hbar\frac{\partial f}{\partial t}=H_{(+)} f.
\end{eqnarray}
Considering that $H_{11}=V+mc^2$, $H_{12}=H_{21}= c\mathcal{P}$ and $H_{22}=V-mc^2$, as well as Exps. (\ref{fi1}) for
$\phi$, we obtain
\begin{eqnarray} \label{14}
\fl H_{(+)}= \cos\phi\ooa V\cos\phi+\sin\phi\ooa V\sin\phi +\sin\phi\ooa (c\mathcal{P})\cos\phi + \cos\phi\ooa
(c\mathcal{P})\sin\phi\nonumber \\ \fl -2 mc^2\sin^2\phi+mc^2 =V+[\cos\phi, V]\cos\phi+[\sin\phi,
V]\sin\phi+c\sqrt{m^2c^2+\mathcal{P}^2}.
\end{eqnarray}
(It should be noted that this expression coincides with the expressions obtained for the transformed FW-Hamiltonian
obtained in closed form for a free particle \cite{Fol}, for a particle in an external vector potential \cite{Cas,Eric} and
for a particle in an external scalar potential commuting with an odd operator \cite{Sil}.)

If we replace the $\phi$ operator in Exp. (\ref{14}) with $\phi_{(-)}$ and take into account Exps. (\ref{fi2}), we obtain
a two-component relativistic Hamiltonian bounded from above:
\begin{eqnarray}\label{20f}
\fl H_{(-)}=\cos\phi\ooa V\cos\phi+\sin\phi\ooa V\sin\phi -\sin\phi\ooa (c\mathcal{P})\cos\phi -\cos\phi\ooa
(c\mathcal{P})\sin\phi\nonumber\\ \fl +2 mc^2\sin^2\phi-mc^2 =V+[\cos\phi, V]\cos\phi+[\sin\phi,
V]\sin\phi-c\sqrt{m^2c^2+\mathcal{P}^2}.
\end{eqnarray}
It is important to emphasize here that expressions (\ref{14}) and (\ref{20f}), obtained in closed analytical form, are
valid for both vector and scalar fields. At the same time, obtaining the FW transformation for a non-zero scalar potential
is a complex task.

To compare Eq. (\ref{3f}) with the Pauli equation, it is necessary to make in (\ref{3f}) the substitution $f=\widetilde{f}
\exp(-mc^2t/\hbar)$. This results in a ``two-component'' equation that is valid for any particle velocity and any
external, vector, or scalar fields:
\begin{eqnarray} \label{3ff}
\fl i\hbar\frac{\partial \widetilde{f}}{\partial t}=\widetilde{H}_{(+)} \widetilde{f};\ppp
\widetilde{H}_{(+)}=H_{(+)}-mc^2.
\end{eqnarray}

\section{A two-component analogue of the probability current density} \label{current}

In this section we introduce the `two-component' analogues of the probability density $w$ and the probability current
density $\vec{J}$ that enter the continuity equation
\begin{eqnarray*}
\fl \frac{\partial w}{\partial t}+\vec{\nabla}\vec{J}=0;\ppp w=\Phi^\dag\Phi+\chi^\dag\chi,\ppp
\vec{J}=c(\Phi^\dag\vec{\sigma}\chi+\chi^\dag\vec{\sigma}\Phi),
\end{eqnarray*}
They are defined through the spinor $f$ as follows (here $G=\cos\phi$):
\begin{eqnarray} \label{15}
\fl w_f=f^\dag f,\ppp \vec{J}_f=c(Gf)^\dag\left(\vec{\sigma} Q+Q\vec{\sigma}\right)Gf.
\end{eqnarray}
For the $\alpha$-th component of $\vec{J}_f$ (summation is assumed over the same indices) we have
\begin{eqnarray*}
\fl (J_f)_\alpha= \frac{1}{2}\left[\left(\sigma_\alpha\frac{\mathcal{P}}{M}Gf\right)^\dag Gf+ (Gf)^\dag\sigma_\alpha
\frac{\mathcal{P}}{M} Gf\right]\\ \fl = \frac{1}{2}\left[\left(\sigma_\alpha \sigma_n\pi_n\frac{1}{M} Gf\right)^\dag (Gf)
+(Gf)^\dag \sigma_\alpha\sigma_n\pi_n \frac{1}{M} Gf\right].
\end{eqnarray*}
Since $\sigma_n\sigma_k=\delta_{nk}\sigma_0+i\varepsilon_{nkl}\sigma_l$, we obtain the decomposition
$\vec{J}_f=\vec{J}_{\vec{\pi}}+\vec{J}_{\vec{\sigma}}$:
\begin{eqnarray*}
\fl \vec{J}_{\vec{\pi}}= \Re\left[(Gf)^\dag  \vec{\pi}\frac{G}{M} f \right],\ppp \vec{J}_{\vec{\sigma}}=\Im\left[(Gf)^\dag
[\vec{\sigma}\times \vec{\pi}]\frac{G}{M} f\right],
\end{eqnarray*}
where $\vec{J}_{\vec{\pi}}$ and $\vec{J}_{\vec{\sigma}}$ are the orbital and spin parts of the total probability current
density. In fact, this is an analogue of the Gordon decomposition \cite{Gor}, that appears in the two-component
formulation.

\section{Free Dirac particle in the two-component formulation} \label{free}

To demonstrate the two-component formulation in `work', we will find the general solution of Eq. (\ref{3ff}) for a free
particle. That is, we need to solve the equation
\begin{eqnarray}\label{335}
\fl i\hbar \frac{\partial \widetilde{f}}{\partial t}=\widetilde{H}_{free} \widetilde{f}; \ppp \widetilde{H}_{free}=
mc^2\left(\sqrt{1+\left(\frac{\vec{p}}{mc}\right)^2}-1\right).
\end{eqnarray}

We will look for its particular solution in the form
\begin{eqnarray*}
\fl \widetilde{f}(\vec{r},t)=f(\vec{r}) e^{-i\varepsilon t/\hbar},
\end{eqnarray*}
where the function $f(\vec{r})$ satisfies the equation
\begin{eqnarray}\label{3355}
\fl mc^2\left(\sqrt{1+\left(\frac{\vec{p}}{mc}\right)^2}-1\right)f(\vec{r})=\epsilon f(\vec{r});
\end{eqnarray}
where $\epsilon>0$. Let $f(\vec{r})=\frac{1}{(2\pi)^{3/2}}\int f(\vec{k}) e^{i\vec{k}\vec{r}} d^3 \vec{k}$, then, up to a
constant spinor, for the Fourier-image of the desired solution we obtain
\begin{eqnarray*}
\fl f(\vec{k})= \delta(\vec{k}-\vec{k}_\epsilon);\ppp |\vec{k}_\epsilon|=\frac{1}{\hbar}
\sqrt{2m\epsilon+\frac{\epsilon^2}{c^2}};
\end{eqnarray*}
the vector $\vec{k}_\epsilon$ depends on the energy $\epsilon$ and two angular parameters that determine its direction;
$|\vec{k}_\epsilon|\neq 0$, since $\epsilon\neq 0$.

Thus, two independent three-parameter stationary solutions of the spinor equation (\ref{335}) can be written in the form
\begin{eqnarray*}
\fl f_1(\vec{r},t;\vec{k}_\epsilon)= \left(\begin{array}{cc} 1 \\
0\end{array}\right) e^{i(\vec{k}_\epsilon\vec{r}-\epsilon t/\hbar)},\ppp f_2(\vec{r},t;\vec{k}_\epsilon)=
\left(\begin{array}{cc} 0 \\
1\end{array}\right) e^{i(\vec{k}_\epsilon\vec{r}-\epsilon t/\hbar)}.
\end{eqnarray*}
They can be rewritten in the form
\begin{eqnarray*}
\fl f_1(\vec{r},t;\vec{k})= \left(\begin{array}{cc} 1 \\
0\end{array}\right) e^{i\left[\vec{k}\vec{r}-\epsilon(\vec{k}) t/\hbar\right]},\ppp f_2(\vec{r},t;\vec{k})= \left(\begin{array}{cc} 0 \\
1\end{array}\right) e^{i\left[\vec{k}\vec{r}-\epsilon(\vec{k}) t/\hbar\right]}.
\end{eqnarray*}
where $\vec{k}$ is an arbitrary non-zero vector, and $\epsilon(\vec{k})=\sqrt{(c\hbar \vec{k})^2+m^2c^4}-mc^2$.

These solutions are obviously not characterized by a certain helicity. At the same time, the helicity operator
$\hat{h}=(\vec{\sigma}\vec{p})/|\vec{p}|$ commutes with the operator $H_{free}$ and, therefore, is an integral of motion.
Therefore, we will look for independent particular solutions of the form $f(\vec{r},t;\vec{k})=\mathcal{C}(\vec{k})
e^{i[\vec{k}\vec{r}-\epsilon(\vec{k}) t/\hbar]}$, which correspond to helicities $+1$ and $-1$. They must satisfy the
equation $\hat{h}f=s f$, where $s=\pm 1$.

Thus, the two-component spinor $\mathcal{C}(\vec{k})=\left(\begin{array}{cc} C_1 \\ C_2\end{array}\right)$ of the desired
solutions must satisfy the equation
\begin{eqnarray*}
\fl \left(\begin{array}{cc} k_z  & k_x-i k_y\\
k_x+ik_y & -k_z\end{array}\right) \left(\begin{array}{cc} C_1 \\ C_2\end{array}\right)=\pm |\vec{k}|\left(\begin{array}{cc} C_1 \\
C_2\end{array}\right)
\end{eqnarray*}
with $|\vec{k}|\neq 0$. Its solutions $\mathcal{C}_{+1}(\vec{k})$ and $\mathcal{C}_{-1}(\vec{k})$, corresponding to
helicities $+1$ and $-1$, can be written in the form
\begin{eqnarray*}
\fl \mathcal{C}_{+1}(\vec{k})= \frac{1}{\sqrt{2 |\vec{k}|(|\vec{k}|+k_z)}}
\left(\begin{array}{cc} |\vec{k}|+k_z  \\
k_x+ik_y\end{array}\right),\ooo \mathcal{C}_{-1}(\vec{k})= \frac{1}{\sqrt{2 |\vec{k}|(|\vec{k}|+k_z)}}
\left(\begin{array}{cc}  k_x-ik_y \\ -|\vec{k}|-k_z
\end{array}\right);
\end{eqnarray*}
they satisfy the conditions
\begin{eqnarray*}
\fl \mathcal{C}_{+1}^\dag(\vec{k})\mathcal{C}_{-1}(\vec{k})=0,\ooo
\mathcal{C}_{+1}^\dag(\vec{k})\mathcal{C}_{+1}(\vec{k})= \mathcal{C}_{-1}^\dag(\vec{k})\mathcal{C}_{-1}(\vec{k})=1.
\end{eqnarray*}
Thus, the stationary solution with the helicity $s$ is $f_s(\vec{r},t;\vec{k})=\mathcal{C}_s(\vec{k})
e^{i[\vec{k}\vec{r}-\epsilon(\vec{k}) t/\hbar]}$ where $s=\pm 1$. And now we can construct the corresponding bispinors --
stationary solutions of the Dirac equation for a free particle.

To do this, we need to find the corresponding large and small components $\Phi$ and $\chi$. For a free particle, the
operators $\cos \phi$ and $\sin \phi$ (see Exps. (\ref{fi1})) have the form:
\begin{eqnarray*}
\fl \cos \phi=W_\Phi(\vec{p}), \ooo W_\Phi(\vec{p})=\frac{1}{\sqrt{2}}\sqrt{1+\frac{1}{\sqrt{1+(\vec{p}/mc)^2}}};\\ \fl
\sin \phi= \hat{h}W_\chi(\vec{p});\ooo W_\chi(\vec{p})= \frac{1}{\sqrt{2}} \sqrt{1-\frac{1}{\sqrt{1+(\vec{p}/mc)^2}}}.
\end{eqnarray*}
Therefore,
\begin{eqnarray*}
\fl \Phi_{s}(\vec{r},t;\vec{k})=(\cos\phi) f_{s}(\vec{r},t;\vec{k})= W_\Phi(\hbar k)f_{s}(\vec{r},t;\vec{k}),\\
\fl \chi_{s}(\vec{r},t;\vec{k})=(\sin\phi) f_{s}(\vec{r},t;\vec{k})= W_\chi(\hbar k)\hat{h} f_{s}(\vec{r},t;\vec{k})=
s\cdot W_\chi(\hbar k) f_{s}(\vec{r},t;\vec{k}).
\end{eqnarray*}
Hence, the corresponding bispinor is
\begin{eqnarray*}
\fl \Psi_{s}(\vec{r},t;\vec{k})=\left(\begin{array}{cc} W_\Phi(\hbar k) f_{s}(\vec{r},t;\vec{k}) \\
s \cdot W_\chi(\hbar k)f_{s}(\vec{r},t;\vec{k})
\end{array}\right)=\left(\begin{array}{cc} W_\Phi(\hbar k) \mathcal{C}_s(\vec{k}) \\
s \cdot W_\chi(\hbar k)\mathcal{C}_s(\vec{k})
\end{array}\right)
e^{i[\vec{k}\vec{r}-\epsilon(\vec{k}) t/\hbar]};
\end{eqnarray*}
recall that $W_\Phi^2(\hbar k)+ W_\chi^2(\hbar k)=1$.

The wave packet normalized to unity, constructed from plane waves with helicity $s$, can now be written as
\begin{eqnarray*}
\fl \Psi_s(\vec{r},t)=\frac{1}{(2\pi)^{3/2}} \int_{\hat{\mathbf{R}}^3} \mathcal{A}_{s}(\vec{k})
\Psi_{s}(\vec{r},t;\vec{k}) d^3\mathbf{k};
\end{eqnarray*}
where the functions $\mathcal{A}_{s}(\vec{k})$ ($s=\pm 1$) satisfy the normalization conditions $\int_{\hat{\mathbf{R}}^3}
\mathcal{A}_{s}^\dag(\vec{k})\mathcal{A}_{s}(\vec{k}) d^3\mathbf{k}=1$. Thus, the unit bispinor describing an arbitrary
state of a free Dirac particle is
\begin{eqnarray*}
\fl \Psi(\vec{r},t)=c_1\Psi_{+1}(\vec{r},t) + c_2\Psi_{-1}(\vec{r},t),
\end{eqnarray*}
where $|c_1|^2+|c_2|^2=1$.

\section{Operators of observables in the space of two-component unit spinors} \label{operators}

Let a self-adjoint Dirac operator in the standard representation have the form
\begin{eqnarray} \label{19}
\fl \mathcal{A}_D=\left(
\begin{array}{cc}
\nu_1 & \nu_2 \\
\nu_2^\dag & \nu_3
\end{array} \right);\ppp \nu_1^\dag=\nu_1,\ooo \nu_3^\dag=\nu_3;
\end{eqnarray}
it is assumed that this operator can be either scalar or vector. As in the case of $H_D$, the self-adjointness condition
$\langle\Psi_1|\mathcal{A}_D\Psi\rangle= \langle \mathcal{A}_D\Psi_1|\Psi\rangle$ for this operator in the space of
four-component unit bispinors (here $\Psi_1$ and $\Psi$ are orbitrary bispinors (see (\ref{1q}))) is reduced, to the
analogous condition $\langle f_1|\mathcal{A}_2f\rangle= \langle \mathcal{A}_2 f_1|f\rangle$ in the space of two-component
unit spinors, where
\begin{eqnarray} \label{20}
\fl \mathcal{A}_f=\cos(\phi) \nu_1\cos(\phi) +\cos(\phi)\nu_2\sin(\phi)+\sin(\phi)\nu_2^\dag\cos(\phi)+\sin(\phi)
\nu_3\sin(\phi).
\end{eqnarray}
Note, in the Foldy-Wouthuysen approach, the transformed operators are represented by four-component matrices, which in
general do not have a block-diagonal form.

It is easy to show that, when $\vec{A}(\vec{r})=0$, the most mysterious operator $\vec{v}=c\vec{\alpha}$, for which
$\nu_1=\nu_3=0$, and $\nu_2=c\vec{\sigma}$, acquires a clear physical meaning in this two-component formulation. Indeed,
when $\vec{A}(\vec{r})=0$, the angular operator $\phi$ depends only on $p^2$, since $\mathcal{P}=(\vec{\sigma}\vec{p})$
and $\mathcal{P}^2=\vec{p}^2$. Hence, $[\nu_2,\phi]=0$, and we obtain
\begin{eqnarray}\label{1O}
\fl \vec{v}_f=\cos\phi\ooa c\vec{\sigma}\sin\phi + \sin\phi\ooa c\vec{\sigma}\cos\phi
=\frac{\vec{\sigma}(\vec{\sigma}\vec{p})+(\vec{\sigma}\vec{p})\vec{\sigma}}{M+\mu}= \frac{\vec{p}}{M+\mu}=\frac{
\vec{p}}{m\sqrt{1+\frac{\vec{p}^2}{m^2c^2}}}.
\end{eqnarray}

\subsection{Two-component operators in the non-relativistic limit} \label{nonrel}

{\bf Dirac operator $H$}: Representing Exp. (\ref{14}) as a series in powers of the `small parameter'
$(\mathcal{P}/mc)^2\ll 1$, we obtain, up to the first order ($N=1$) of smallness in this operator, the well-known result
(see \cite{Ber}) --
\begin{eqnarray*}
\fl
H_{(+)}^{(1)}=\frac{\mathcal{P}^2}{2m}+V+\frac{1}{4m^2c^2}\left[\mathcal{P}V\mathcal{P}-\frac{1}{2}\left(\mathcal{P}^2V+
V\mathcal{P}^2\right)\right]-\frac{\mathcal{P}^4}{8m^3c^2}.
\end{eqnarray*}
Since
\begin{eqnarray*}
\fl \mathcal{P}^2=\pi^2-\frac{e\hbar}{c}(\vec{\sigma}\vec{H});\ooo
\mathcal{P}V\mathcal{P}-\frac{1}{2}\left(\mathcal{P}^2V+V\mathcal{P}^2\right)=\frac{1}{2}[\mathcal{P},[V,\mathcal{P}]]\\
\fl =-\frac{1}{2} e\hbar^2 \mathrm{div}\vec{E}- e\hbar \left(\vec{\sigma}[\vec{E}\times\vec{\pi}]\right);\ooo
\vec{H}=[\vec{\nabla}\times\vec{A}],\ooo \vec{E}=-\vec{\nabla}\phi,
\end{eqnarray*}
we finally obtain the Pauli Hamiltonian with the first relativistic correction:
\begin{eqnarray} \label{14a}
\fl H_{(+)}^{(1)}=\frac{1}{2m}\left[\pi^2-\frac{e\hbar}{c}(\vec{\sigma}\vec{H})\right]+V-\frac{1}{8m^3c^2}\left[\pi^2
-\frac{e\hbar}{c}(\vec{\sigma}\vec{H})\right]^2 \nonumber\\
\fl -\frac{e\hbar^2}{8m^2c^2}\mathrm{div}\vec{E}-\frac{e\hbar}{4m^2c^2}\left(\vec{\sigma}[\vec{E}\times\vec{\pi}]\right)
\end{eqnarray}
The third term in Exp. (\ref{14a}) is is the kinetic energy correction (see, e.g., \cite{Thal}). The fourth term is is
known as the Darwin term. The last term is a generalization, onto the case $\vec{A}\neq 0$ , of the spin orbit coupling
term which joins the terms
\begin{eqnarray*}
\fl -\frac{e\hbar}{2m^2c^2}\left(\vec{\sigma}[\vec{E}\times\vec{p}]\right),\ppp
\frac{e\hbar}{4m^2c^2}\left(\vec{\sigma}[\vec{E}\times\vec{p}]\right)
\end{eqnarray*}
describing the spin-orbit interaction and the Thomas precession, respectively.

{\bf The probability current densities $\vec{J}_{\vec{\pi}}$ and $\vec{J}_{\vec{\sigma}}$}: For $N=0$
\begin{eqnarray} \label{17}
\fl \vec{J}_{\vec{\pi}}^{(0)}= \frac{1}{m} \Re\left[f^\dag (\vec{\pi} f)\right]=\frac{\hbar}{2m}\left[(\vec{\nabla}f^\dag)
f-f^\dag \vec{\nabla}f\right]-\frac{e}{mc}\vec{A}f^\dag f, \nonumber\\\fl  \vec{J}_{\vec{\sigma}}^{(0)}= \frac{1}{m}
\Im\left(f^\dag [\vec{\sigma}\times\vec{\pi}] f\right)=\frac{\hbar}{2m}\left[\vec{\nabla}\times
f^\dag\vec{\sigma}f\right].
\end{eqnarray}
For $N=1$, with the first relativistic corrections, we have
$\vec{J}_{\vec{\pi}}^{(1)}=\vec{J}_{\vec{\pi}}^{(0)}+\vec{\Delta}_{\vec{\pi}}^{(1)}$,
$\vec{J}_{\vec{\sigma}}^{(1)}=\vec{J}_{\vec{\sigma}}^{(0)}+\vec{\Delta}_{\vec{\sigma}}^{(1)}$, where
\begin{eqnarray*}
\fl \vec{\Delta}_{\vec{\pi}}^{(1)}= - \frac{\Re\left[(\mathcal{P}^2 f)^\dag \vec{\pi} f  + 3f^\dag \vec{\pi} \mathcal{P}^2
f\right]}{8 m^3c^2},\ooo \vec{\Delta}_{\vec{\sigma}}^{(1)}= - \frac{\Im\left[(\mathcal{P}^2 f)^\dag
[\vec{\sigma}\times\vec{\pi}] f  + 3f^\dag [\vec{\sigma}\times\vec{\pi}] \mathcal{P}^2 f\right]}{8 m^3c^2}
\end{eqnarray*}
(note that $\mathcal{P}^2=\pi^2-\frac{e\hbar}{c}(\vec{\sigma}\vec{H})$).

{\bf Even Hermitian operators $\mathcal{E}$ -- the position, momentum and angular momentum operators}: For $N=1$ we have
\begin{eqnarray} \label{402}
\fl \mathcal{E}_f^{(1)}=\nu_1+ \frac{W}{8m^2c^2};\ppp W=2\mathcal{P} \nu_3 \mathcal{P} -\mathcal{P}^2\nu_1-\nu_1
\mathcal{P}^2.
\end{eqnarray}
If $\nu_3=\nu_1=\nu$ and, in addition, the operator $\nu$ does not contain Pauli matrices, then
\begin{eqnarray}\label{421}
\fl W=-i\varepsilon_{nml}\sigma_l\{\pi_m,[\nu,\pi_n]\}+[\pi_n,[\nu,\pi_n]];\nonumber\\ \fl
\{\pi_m,[\nu,\pi_n]\}=a^2\{A_m,[\nu,A_n]\}+a\{p_m,[\nu,A_n]\}+a\{A_m,[\nu,p_n]\}+\{p_m,[\nu,p_n]\},\nonumber\\ \fl
\left[\pi_n,[\nu,\pi_n]
\right]=a^2\left[A_n,[\nu,A_n]\right]+a\left[p_n,[\nu,A_n]\right]+a\left[A_n,[\nu,p_n]\right]+\left[p_n,[\nu,p_n]\right];
\end{eqnarray}
hereinafter, the brackets $\{ , \}$ denote the anticommutator; $a=-e/c$.

In particular, {\it for the position operator}, when $\nu=x_\alpha$ ($\alpha=1,2,3$) we have
$[x_\alpha,p_n]=i\hbar\delta_{\alpha n}$ and $[x_\alpha,A_n]=0$; here $\delta_{\alpha n}$ is the Kronecker delta. Thus,
\begin{eqnarray*}
\fl \{\pi_m,[x_\alpha,\pi_n]\}=2i\hbar \delta_{\alpha n}\pi_m,\ooo \left[\pi_n,[x_\alpha,\pi_n] \right]=0;\ooo
W=2\hbar\varepsilon_{\alpha ml}\pi_m \sigma_l=2\hbar\ooa[\vec{\pi}\times\vec{\sigma}]_\alpha.
\end{eqnarray*}
That is
\begin{eqnarray} \label{3x}
\fl \vec{r}_f^{(1)}=\vec{r}+
\frac{\hbar}{4m^2c^2}[\vec{\pi}\times\vec{\sigma}]=\vec{r}+\frac{1}{2m^2c^2}[\vec{\pi}\times\vec{S}];\ppp
\vec{S}=\frac{\hbar}{2}\vec{\sigma}.
\end{eqnarray}
Thus, the components $x_\alpha$ and $x_\beta$ $(\alpha\neq\beta)$ of the vector operator $\vec{r}_f$ commute with each
other only when $N=0$. But already when $N=1$
\begin{eqnarray} \label{4x}
\fl [x_\alpha^{(1)},x_\beta^{(1)}]=- \frac{i\hbar^2}{2m^2c^2}\varepsilon_{\alpha\beta\gamma}\sigma_\gamma=-
\frac{i}{2}\lambda_C^2 \varepsilon_{\alpha\beta\gamma}\sigma_\gamma;\ppp \lambda_C=\frac{\hbar}{mc}
\end{eqnarray}
(for spin-induced non-commutativity of the position operator in relativistic quantum mechanics, see \cite{Der} and
references therein).

{\it For the momentum operator}, when $\nu=p_\alpha$ $(\alpha=1,2,3)$, we have $[p_\alpha,p_n]=0$ and
$[p_\alpha,A_n]=-i\hbar
\partial A_n/\partial x_\alpha$:
\begin{eqnarray*}
\fl \{\pi_m,[p_\alpha,\pi_n]\}=-2i\hbar a^2 A_m\frac{\partial A_n}{\partial x_\alpha}- \hbar^2
a\frac{\partial^2A_n}{\partial x_m\partial x_\alpha}- 2i\hbar a\frac{\partial A_n}{\partial x_\alpha} p_m,,\\
\fl \left[\pi_n,[p_\alpha,\pi_n] \right]=-\hbar^2 a\frac{\partial^2 A_n}{\partial x_n\partial x_\alpha};\ooo W=-i\hbar^2
a\frac{\partial (\vec{\sigma}\vec{H})}{\partial x_\alpha}-2a\hbar \left(\frac{\partial \vec{A}}{\partial
x_\alpha}[\vec{\pi}\times\vec{\sigma}]\right)-\hbar^2 a\frac{\partial (\vec{\nabla}\vec{A})}{\partial x_\alpha}.
\end{eqnarray*}
That is
\begin{eqnarray} \label{1p}
\fl (p_\alpha)_f^{(1)}=p_\alpha+ \frac{i\hbar^2 e}{8m^2c^3}\frac{\partial (\vec{\sigma}\vec{H})}{\partial x_\alpha}
+\frac{\hbar^2 e}{8m^2c^3}\frac{\partial (\vec{\nabla}\vec{A})}{\partial x_\alpha}+ \frac{\hbar
e}{4m^2c^3}\left(\frac{\partial \vec{A}}{\partial x_\alpha}[\vec{\pi}\times\vec{\sigma}]\right).
\end{eqnarray}
From this expression it follows that if $\vec{A}=0$, then $\vec{p}_f^{(1)}=\vec{p}$.

{\it For the angular momentum operator}, when
$\nu=L_\alpha=[\vec{r}\times\vec{p}]_\alpha=\varepsilon_{\alpha\beta\gamma}x_\beta p_\gamma$,
\begin{eqnarray*}
\fl [L_\alpha,p_n]=i\hbar \varepsilon_{\alpha\beta\gamma} p_\gamma\delta_{\beta n}=i\hbar \varepsilon_{\alpha n\gamma}
p_\gamma,\ppp [L_\alpha,A_n]=-i\hbar \varepsilon_{\alpha\beta\gamma}x_\beta \frac{\partial A_n}{\partial x_\gamma}.
\end{eqnarray*}
Substituting these expressions into (\ref{421}) for $\nu=L_\alpha$, we can obtain the desired expression $W$ for
$(L_\alpha)_ f^{(1)}$. In the general case, the final expression is quite cumbersome, but in the case of A=0 it is
significantly simplified:
\begin{eqnarray*}
\fl \vec{L}_f^{(1)}= \vec{L}+ \frac{1}{2m^2c^2}\left(\vec{S}p^2-\vec{p}(\vec{S}\vec{p})\right).
\end{eqnarray*}

{\bf For the spin operator}, when $\nu_1=\nu_3=\vec{S}=\frac{\hbar}{2}\vec{\sigma}$, Eps. (\ref{421}) are not applicable.
Now we have
\begin{eqnarray*}
\fl \vec{S}_f^{(1)}=\vec{S}+\frac{1}{8m^2c^2}[\mathcal{P},[\vec{S},\mathcal{P}]];\ooo
[\vec{S},\mathcal{P}]=i\hbar[\vec{\pi}\times\vec{\sigma}].
\end{eqnarray*}
Therefore,
\begin{eqnarray} \label{24}
\fl \vec{S}_f^{(1)}=\vec{S}+\frac{1}{4m^2c^2}\left[\vec{\pi}(\vec{S}\vec{\pi})+(\vec{S}\vec{\pi})\vec{\pi}-2\vec{S}\pi^2
-2i[\vec{\pi}\times\vec{\pi}]\right];
\end{eqnarray}
note that $[\vec{\pi}\times\vec{\pi}]=i\frac{e\hbar}{c}\vec{H}$. If $\vec{A}=0$, then
$\vec{L}_f^{(1)}+\vec{S}_f^{(1)}=\vec{L}+\vec{S}$. Ultimately, this is explained by the fact that
$[\vec{L}+\vec{S},\mathcal{P}]=0$ for $\vec{A}=0$, although $[\vec{L},\mathcal{P}]\neq 0$ and $[\vec{S},\mathcal{P}]\neq
0$.

{\bf For the operator $\beta$}, which appears in the parity operator $P=\beta p^{(0)}$, where $p^{(0)}$ is the ``orbital
parity'', $\nu_1=-\nu_3=1$ and $\nu_2=0$. Hence
\begin{eqnarray} \label{424}
\fl \beta_f=\cos(2\phi);\ppp \beta_f^{(1)}=1-\frac{3\mathcal{P}^2}{8m^2c^2}.
\end{eqnarray}

{\bf For the velocity operator $\vec{v}= c\vec{\alpha}$}, when $\nu_1=\nu_3=0$ and $\nu_2=c\vec{\sigma}$ in (\ref{19}),
for $N=1$ we have
\begin{eqnarray*}
\fl \vec{v}_f^{(1)}=\frac{1}{2m}(\vec{\sigma} \mathcal{P}+\mathcal{P} \vec{\sigma}) -\frac{1}{16m^3c^2}\left(3\vec{\sigma}
\mathcal{P}^3+3 \mathcal{P}^3\vec{\sigma}+\mathcal{P}^2\vec{\sigma} \mathcal{P}+\mathcal{P}\vec{\sigma}
\mathcal{P}^2\right).
\end{eqnarray*}
And since $\{\mathcal{P},\vec{\sigma}\}\equiv\vec{\sigma} \mathcal{P}+\mathcal{P} \vec{\sigma}=2\vec{\pi}$, we have
\begin{eqnarray} \label{25}
\fl \vec{v}_f^{(1)}=\frac{\vec{\pi}}{m} -\frac{1}{8m^3c^2}\left(3\vec{\pi} \mathcal{P}^2+3
\mathcal{P}^2\vec{\pi}-2\mathcal{P}\vec{\pi} \mathcal{P}\right).
\end{eqnarray}

{\bf For the spirality operator $\gamma^5$}, when $\nu_1=\nu_3=0$ and $\nu_2=\sigma_0$, we have
\begin{eqnarray} \label{429}
\fl \gamma^5_f=\sin(2\phi)=\frac{\xi}{\sqrt{1+\xi^2}};\ppp (\gamma^5_f)^{(1)}=\xi=\frac{\mathcal{P}}{mc}.
\end{eqnarray}

\subsection{Two-component operators in the strict ultrarelativistic limit} \label{ultrarel}

The study of this limit is relevant for neutrinos and antineutrinos, and we will consider this limit for a free Dirac
particle. All expressions for the two-component operators are now series in powers of the small parameter
$\rho=mc/|\vec{p}|$. But we will retain only the leading terms of the expansion in this parameter. In particular, in the
strict ultrarelativistic limit, the operator
\begin{eqnarray}\label{neu}
\fl H_{free}= c|\vec{p}|\left(\sqrt{1+\rho^2}-\rho\right)
\end{eqnarray}
is equal to $c|\vec{p}|$, and the stationary equation for the spinor $f$ is written as
\begin{eqnarray}\label{411}
\fl c|\vec{p}|f=\epsilon f
\end{eqnarray}
where $\epsilon\geq 0$. And since
\begin{eqnarray*}
\fl |\vec{p}|=|\mathcal{P}|=\left\{\begin{array}{cc}
\mathcal{P}\ppp\mbox{when}\ppp \mathcal{P}>0, \\
-\mathcal{P}\ppp\mbox{when}\ppp \mathcal{P}<0
\end{array}\right.,
\end{eqnarray*}
Eq. (\ref{411}) `splits' into two independent ones:
\begin{eqnarray}\label{412}
\fl \mbox{(a)}\ooo \left[c(\vec{\sigma}\vec{p})+\epsilon\right] f_a=0;\ppp \mbox{and}\ppp \mbox{(b)}\ooo
\left[c(\vec{\sigma}\vec{p})-\epsilon\right] f_b=0,
\end{eqnarray}
where the first (second) descibes the states with the negative (positive) helicity. That is, as in \cite{Cin,Boz}, we have
obtained a form of Hamiltonian such that states with positive and negative helicity are described by two independent
two-component equations. For these strictly ``longitudinal polarizations'', all expressions for two-component operators in
this limit must be represented taking into account the equality $[\vec{\sigma}\times\vec{p}]=0$.

{\bf For the probability current density} we have
\begin{eqnarray}\label{412a}
\fl \vec{J}_{\vec{\pi}}^{as}=c\Re\left(f^\dag \frac{\vec{p}}{|\vec{p}|} f\right),\ppp
\vec{J}_{\vec{\sigma}}^{as}=c\Im\left(f^\dag \frac{[\vec{\sigma}\times\vec{p}]}{|\vec{p}|} f\right)=0,
\end{eqnarray}
that is, the spin current is zero in this limit.

{\bf For the operators $(\cos(\phi))_{as}$ and $(\sin(\phi))_{as}$}, which relate the large and small components to the
normalized spinor $f$, we have
\begin{eqnarray}\label{410}
\fl \cos\phi_{as}=\frac{1}{\sqrt{2}},\ppp \sin\phi_{as}=\frac{1}{\sqrt{2}}\frac{\mathcal{P}}{|\mathcal{P}|}=
\frac{1}{\sqrt{2}}\frac{(\vec{\sigma}\vec{p})}{|\vec{p}|};
\end{eqnarray}
as is seen, $\sin\phi_{as}$ coincides up to the factor $1/\sqrt{2}$ with the helicity operator.

{\bf For any even operator $\mathcal{E}$}, for both classes of states, we have
\begin{eqnarray} \label{404}
\fl \mathcal{E}_f^{as}= \frac{1}{2}\left(\nu_1+ \frac{\mathcal{P}}{|\vec{p}|}
\nu_3\frac{\mathcal{P}}{|\vec{p}|}\right)=\nu_1+\nu_3.
\end{eqnarray}
In particular,
\begin{eqnarray}\label{405}
\fl \vec{r}_f^{as}=\vec{r},\ppp \vec{p}_f^{as}=\vec{p},\ppp \vec{L}_f^{as}=\vec{L},\ppp \vec{S}_f^{as}=\vec{S}\equiv
\vec{p}\frac{(\vec{S}\vec{p})}{|\vec{p}|^2}.
\end{eqnarray}
{\bf For the operator $\beta$} (and hence the relative intrinsic parity operator $P=\beta P^{(0)}$ where $P^{(0)}$ is the
`orbital parity' operator, it is equal to zero in this limit, since $\nu_1+\nu_3=0$.

{\bf For any odd operator $\mathcal{O}$}
\begin{eqnarray*}
\fl \mathcal{O}_f^{as}=
\frac{1}{2}\left(\nu_2\frac{\mathcal{P}}{|\vec{p}|}+\frac{\mathcal{P}}{|\vec{p}|}\nu_2^\dag\right).
\end{eqnarray*}
In particular, {\bf for the velocity operator}, since $\vec{\sigma}\mathcal{P}+\mathcal{P}\vec{\sigma}=2\vec{p}$, we have
\begin{eqnarray}\label{406}
\fl \vec{v}_f^{as}= c\frac{\vec{p}}{|\vec{p}|}.
\end{eqnarray}
As is seen, $|\vec{v}_f^{as}|=c$ in the ultrarelativistic limit.

{\bf The spirality operator $\gamma^5$}: its two-component analogue coincides in this limit with the helicity operator
$\hat{h}$:
\begin{eqnarray} \label{429a}
\fl \left(\gamma^5_f\right)_{as}=\frac{\mathcal{P}}{|\mathcal{P}|}=\frac{(\vec{\sigma}\vec{p})}{|\vec{p}|}.
\end{eqnarray}

\section{On the FW formalism with ``the wave function reduction condition''} \label{Nezn}

Let us compare our approach with the FW formalism and clarify the role of the ``wave function reduction condition''. We
begin with the fact that the operator $H_{(+)}$ (see (\ref{14})) coincides with the corresponding FW Hamiltonians in the
case of a free particle, as well as a particle in external fields described by a vector potential and a scalar potential
commuting with an odd operator. It is easy to verify that in all these cases, the unitary operator $U$, which transforms
the Dirac bispinor $\Psi$ into the FW bispinor $\Psi_{FW}=U\Psi_D$, can be written in terms of the operator $\phi$ (see
(\ref{fi1})):
\begin{eqnarray} \label{460}
\fl U=\left(
\begin{array}{cc}
\cos\phi & \sin\phi \\
-\sin\phi & \cos\phi
\end{array} \right).
\end{eqnarray}
In the general case, both components of the transformed bispinor are nonzero:
\begin{eqnarray*}
\fl \Psi_{FW}=\left(
\begin{array}{cc}
\cos\phi & \sin\phi \\
-\sin\phi & \cos\phi
\end{array} \right) \left(
\begin{array}{cc}
\Phi \\ \chi
\end{array} \right)= \left(
\begin{array}{cc}
\cos\phi\ooa\Phi+\sin\phi\ooa\chi \\ -\sin\phi\ooa\Phi+\cos\phi\ooa\chi
\end{array} \right).
\end{eqnarray*}
And all our criticisms of the FW approach presented in the introduction remain valid.

However, the situation changes cardinally if the FW transformation is considered in the space of bispinors whose large and
small components are associated with the same two-component unit spinor (see (\ref{888})). Namely, for the Dirac bispinors
\begin{eqnarray} \label{461}
\fl \Psi_D^{(+)}= \left(
\begin{array}{cc}
\cos\phi\ooa f \\ \sin\phi\ooa f \end{array} \right),\ppp \Psi_D^{(-)}= \left(
\begin{array}{cc}
\cos\phi_{(-)}\ooa f \\ \sin\phi_{(-)}\ooa f \end{array} \right)=\left(
\begin{array}{cc}
-\sin\phi\ooa f \\ \cos\phi\ooa f \end{array} \right)
\end{eqnarray}
we have
\begin{eqnarray*}
\fl \Psi_{FW}^{(+)}= \left(
\begin{array}{cc}
f \\ 0 \end{array} \right),\ppp \Psi_{FW}^{(-)}= \left(
\begin{array}{cc}
0 \\ f \end{array} \right).
\end{eqnarray*}
That is, the Dirac bispinors (\ref{461}) satisfy the wave function reduction condition introduced in \cite{Nez}.

A more intriguing situation arises for an external scalar potential depending on $\mathbf{r}$, which does not commute with
the momentum operator. At first glance, in this case the unitary operator (\ref{460}) does not diagonalize the Dirac
Hamiltonian. But let us compare our two-component operator $H_{(+)}$ with the FW Hamiltonian $H_{FW}^{(+)}$ obtained in
\cite{Nez1} in the Weyl representation for an arbitrary scalar potential $V(\mathbf{r})=eA^0(\mathbf{r})$ for the positive
energy $\varepsilon=|E|>0$; note that $\hbar=c=1$ in \cite{Nez1}.

According to (9) in \cite{Nez1}, in order to rewrite all expressions of \cite{Nez1} in the standard representation, we
have to use the substitutions $m\leftrightarrow \mathcal{P}$ and $\beta\leftrightarrow \gamma_5$; in the designations of
the corresponding quantities we will add the ``wave'' at the top: $U_{FW}^{(+)}\rightarrow \widetilde{U}_{FW}^{(+)}$,
$H_{FW}^{(+)}\rightarrow \widetilde{H}_{FW}^{(+)}$ and $A_{(+)}\rightarrow \widetilde{A}_{(+)}$. Thus, the unitary
operator $U_{FW}^{(+)}$ and the FW Hamiltonian \cite{Nez1} corresponding to the Dirac Hamiltonian in the standard
representation are equal to
\begin{eqnarray*}
\fl \widetilde{U}_{FW}^{(+)}=\widetilde{A}_{(+)} \left(1+\frac{1}{\varepsilon+m-eA^0}\beta\gamma_5\mathcal{P}\right),\ooo
\widetilde{A}_{(+)} = \left[1+\frac{\mathcal{P}^2}{(\varepsilon+m-eA^0)^2}\right]^{-1/2},
\end{eqnarray*}
\begin{eqnarray*}
\fl \widetilde{H}_{FW}^{(+)}=\widetilde{A}_{(+)}\left[m+\frac{2\mathcal{P}^2}{\varepsilon+m-eA^0}-
\frac{\mathcal{P}^2}{\varepsilon+ m-eA^0}\frac{1}{\varepsilon+m-eA^0}\right]\widetilde{A}_{(+)}\\ \fl +
\widetilde{A}_{(+)}\left[eA^0+\frac{\mathcal{P}}{\varepsilon+
m-eA^0}eA^0\frac{\mathcal{P}}{\varepsilon+m-eA^0}\right]\widetilde{A}_{(+)}.
\end{eqnarray*}
As is seen, the operator $\widetilde{H}_{FW}^{(+)}$, unlike $H_{(+)}$, depends nonlinearly on the scalar potential $eA^0$.
And, at first glance, these two approaches lead to different two-component relativistic Hamiltonians for a particle moving
under the action of a scalar potential.

However, there is one interesting point. The key difference between FW approach \cite{Nez1} and ours is that in the former
the operator $\varepsilon-eA^0$ does not commute with $\mathcal{P}$, whereas in our approach, the corresponding operator
$D_t^{(+)}-V$ satisfies equation (\ref{2t}) and, as a consequence, commutes with $\mathcal{P}$. Thus, if we take into
account that $D_t^{(+)}-V=\sqrt{m^2+\mathcal{P}^2}$ (see (\ref{3t})), then
\begin{eqnarray*}
\fl \widetilde{A}_{(+)} = \cos\phi,\ooo \frac{1}{\varepsilon +m-eA^0}\mathcal{P}=\tan\phi.
\end{eqnarray*}
It is easy to show that in this case $\widetilde{U}_{FW}^{(+)}=U$ and $\widetilde{H}_{FW}^{(+)}=H_{(+)}$. This gives
grounds to believe that in the exact FW formalism with the ``wave function reduction condition'' the operator
$\varepsilon-eA^0$ should commute with $\mathcal{P}$, as in our approach.

\section{Conclusion}

Dirac theory for a particle in static external fields is formulated as a two-component single-particle quantum theory, in
which the particle states are described by unit two-component spinors. The original ``four-component'' Dirac Hamiltonian
is split into two operators acting in the space of unit two-component spinors: one is bounded below and is considered a
``two-component'' Hamiltonian describing the Dirac particle, while the other is bounded above and can be ignored as having
no physical meaning when external fields are sufficiently weak (an investigation of its role in the case of strong fields
is planned for another paper). Expressions for the two-component analogs of the Dirac Hamiltonian and other operators of
the standard representation are obtained in closed analytical form, which are then investigated in the nonrelativistic and
ultrarelativistic limits. In addition, a general solution of the Dirac equation describing a free particle, found within
the framework of this approach, is presented.

\section*{Data Availability}

No datasets were generated or analysed during the current study.

\section*{Funding}

No external funding was received for this research.

\end{document}